\documentclass[a4paper]{jpconf}
\usepackage{graphicx}
\usepackage{epsfig,cite,colordvi}
\usepackage{fmtcount}

\begin{document}
\title{Fast-timing measurements in $^{95,96}$Mo}

\author{
  S~Kisyov$^{1}$,   {S~Lalkovski}$^{1}$,   {N~M\v{a}rginean}$^{2}$,   
{D~Bucurescu}$^{2}$,   {L~Atanasova}$^{3}$,   {D~Balabanski}$^{3}$,  
{Gh~Cata-Danil}$^{2}$,   {I~Cata-Danil}$^{2}$,  {D~Deleanu}$^{2}$,   
{P~Detistov}$^{3}$,  {D~Filipescu}$^{2}$,   {D~Ghita}$^{2}$,   
{T~Glodariu}$^{2}$,   {R~M\v{a}rginean}$^{2}$,   {C~Mihai}$^{2}$, 
{A~Negret}$^{2}$, {S~Pascu}$^{2}$, {T~Sava}$^{2}$,   {L~Stroe}$^{2}$,   
{G~Suliman}$^{2}$,   {N V Zamfir}$^{2}$ and  {M~Zhekova}$^{1}$
}

\vspace{0.3cm}

\address{
$^1$Faculty of Physics, University of Sofia "St.~Kliment Ohridski", Sofia, 
Bulgaria \\
$^2$National Institute for Physics and Nuclear Engineering "Horia Hulubei", 
Magurele,
Romania \\
$^3$Institute for Nuclear Research and Nuclear Energy, Bulgarian
Academy of Science, Sofia, Bulgaria \\
}

\ead{stanimir.kisyov@phys.uni-sofia.bg}

\begin{abstract}
Half-lives of the $19/2_1^+$ and $21/2_1^+$ states in $^{95}$Mo and of the 
$8_1^+$ and $10_1^+$ states in $^{96}$Mo were measured. Matrix elements for yrast 
transitions in $^{95}$Mo and $^{96}$Mo are discussed.
\end{abstract}

\section{Introduction}
There are twenty six molibdenum isotopes on the Segr\'e chart, studied by means
of $\gamma$-ray spectroscopy \cite{nndc}. They span a wide region, where a 
variety of shapes are observed, from the spherical $^{90}$Mo to one of the 
most deformed nuclei in the $A\approx 110$ mass region. Of particular interest 
are the $^{95,96}$Mo nuclei, which are placed in a region where an offset of 
the quadrupole deformation takes place \cite{Ca00}. Also, the two nuclei are 
placed in the third oscillator shell in protons and at the beginning of the 
fourth oscillator shell in neutrons, where high-$j$ and low-$j$ single 
particle orbits with $\Delta j=3$ enhance the effects of octupole collectivity 
\cite{La07}. Being on the edge between the single-particle and collective 
modes, the two nuclei represent an excellent laboratory, where different 
approaches can be tested. In this respect, of particular interest are the 
nuclear lifetimes of excited states which are directly related to transition matrix elements and hence 
are sensitive to the underlying structure.

\section{Experimental Set Up and Data Analysis}
Fast-timing measurements were performed in $^{95,96}$Mo. The nuclei were 
produced in fusion-evaporation reactions, performed at the NIPNE tandem 
accelerator. A beam of $^{18}$O was accelerated up 
to 62 MeV and focused on a $^{82}$Se target with a thickness of 5 mg/cm$^{2}$ 
on a 2 mg/cm$^{2}$ thick Au backing. The cross section for the 
$^{82}$Se($^{18}$O, 5n$\gamma$)$^{95}$Mo channel was estimated as 400 mb, while that for the 
$^{82}$Se($^{18}$O, 4n$\gamma$)$^{96}$Mo channel as about 100 mb. 
The beam intensity was of the order of 20 pnA. 
\begin{figure}[t]
\begin{minipage}{24pc}
\begin{center}
\rotatebox{-90}{\scalebox{0.35}[0.35]{\includegraphics{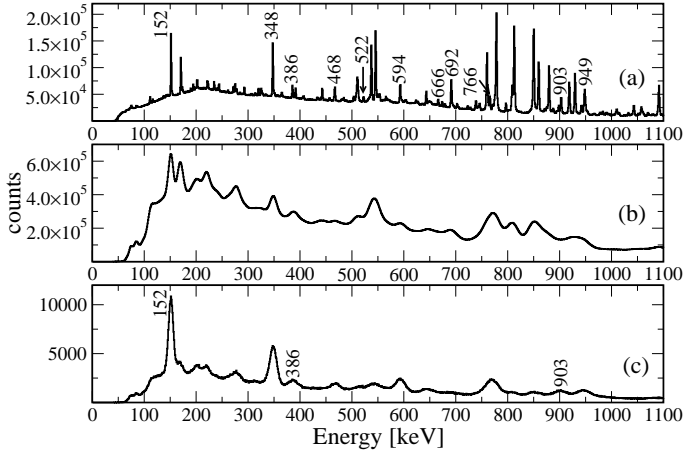}}}
\caption{$^{95}$Mo energy spectra. HPGe total projection (a),   
LaBr$_3$:Ce total projection (b), LaBr$_{3}$:Ce energy spectrum 
gated on 692-keV transition with HPGe detectors (c).}
\label{ene}
\end{center}
\end{minipage}\hspace{1pc}
\begin{minipage}{13pc}
\begin{center}
\scalebox{0.31}[0.31]{\includegraphics{Fig2.epsi}}
\caption{Partial level scheme of $^{95}$Mo.}
\label{shf}
\end{center}
\end{minipage}\hspace{2pc}
\end{figure}

The gamma-ray detector system consisted of eight LaBr$_{3}$:Ce detectors and 
eight HPGe detectors, working in coincidence\cite{Nicu, Kis}. 
The LaBr$_3$:Ce detectors issue positive dynode and negative anode signals.
The dynode signals were used for energy measurements while the anode 
signals were used for timing. The time signals were shaped by Constant 
Fraction Discriminators (CFDs) and sent to Time-to-Amplitude-Converters 
(TACs) operating in a common STOP mode. The energy and time signals were
digitized and the data were stored in event files, collected for approximately 
two hours each. The system was triggered by coincidences between two LaBr$_3$:Ce
and one HPGe detector. The data were analyzed with GASPware and RadWare 
\cite{Rad} packages. 
 
Three dimensional matrices were sorted with LaBr$_{3}$:Ce energy on two of 
the axes and a relative time, defined as $T_{1,2}=t\pm (t_1-t_2)$, on the third 
axis. Here, $t_1$ is the moment of interaction of the preceding in the time 
$\gamma$-ray and $t_2$ is the moment at which the delayed $\gamma$-ray was 
detected. $t$ is an arbitrary offset. If the two $\gamma$-rays feed and 
de-excite a particular nuclear level and $E_{\gamma 1}$ denotes the energy of 
the feeding transition, while $E_{\gamma 2}$ is the energy of the delayed 
transition, then for each event two matrix elements 
$(E_{\gamma 1},E_{\gamma 2}, T_1)$
and $(E_{\gamma 2},E_{\gamma 1}, T_2)$ were incremented. Then, from the time 
projected $(E_{\gamma 1},E_{\gamma 2})$ and $(E_{\gamma 2},E_{\gamma 1})$ 
two-dimensional gates, two time spectra, symmetric with respect to the arbitrary offset $t$, 
were obtained. If the lifetime of the level of interest is 
of the order of the detectors time resolution, the centroids 
$C_D=<t>= \int tD(t)dt /\int D(t)dt$ of the time distributions will be 
displaced by 2$\tau$ one from the other, where $\tau$ is the lifetime of the 
level.  This procedure is known as the centroid shift method \cite{Andr , JMR}.
 It should be noted that the two centroids will overlap in the cases where 
$E_{\gamma 1}$ and $E_{\gamma 2}$ feed and de-excite nuclear level with a 
lifetime shorter than the binning of the converter, which was of 6 ps/channel. For lifetimes longer than the detector time 
resolution the slope method was used to determine level lifetimes. To reduce 
the background in the LaBr$_3$:Ce spectra, the $(E_\gamma ,E_\gamma, T)$ 
matrices were incremented after an $(E_\gamma,T)$ gate was implied on any of 
the HPGe detectors.

Sample energy spectra, obtained with HPGe and LaBr$_3$:Ce detectors, 
are presented in Fig.~\ref{ene}. Fig.\ref{ene}(a)
shows a total energy projection for all HPGe detectors. Peaks 
corresponding to transitions in $^{95}$Mo are denoted with the transition
energies. Fig~\ref{ene}(b) presents the total energy projection obtained 
with all LaBr$_3$:Ce detectors. A LaBr$_3$:Ce spectrum, gated on the 692-keV 
transition in $^{95}$Mo in the HPGe detectors is shown on Fig.~\ref{ene}(c).

The partial level scheme of $^{95}$Mo presented on Fig.~\ref{shf} is based on the 
coincidence measurements performed in the present study and is consistent
with the level scheme presented in ref.~\cite{YHZ}. 
Nuclear level spins and parities J$^{\pi}$, level energies $E_{level}$, $\gamma$-ray energies $E_\gamma$, 
intensities $I_\gamma$ and multipolarities $\lambda$M along with the half-lives 
$T_{1/2}$ of the respective levels in $^{95}$Mo are listed in Table~\ref{tab}. 
Gamma-ray intensities for the strongest yrast transitions were deduced from 
a spectrum gated by the 949-keV transition and normalized to the 
594-keV transition. In the cases where a particular level decays by more 
transitions, spectra gated on the feeding transition were used to normalize
the intensity of the weaker transition to the intensity of the strongest 
transition. Energy and intensity balance was performed.

\begin{figure}[t]
\begin{minipage}{18pc}
\begin{center}
\rotatebox{-90}{\scalebox{0.30}[0.30]{\includegraphics{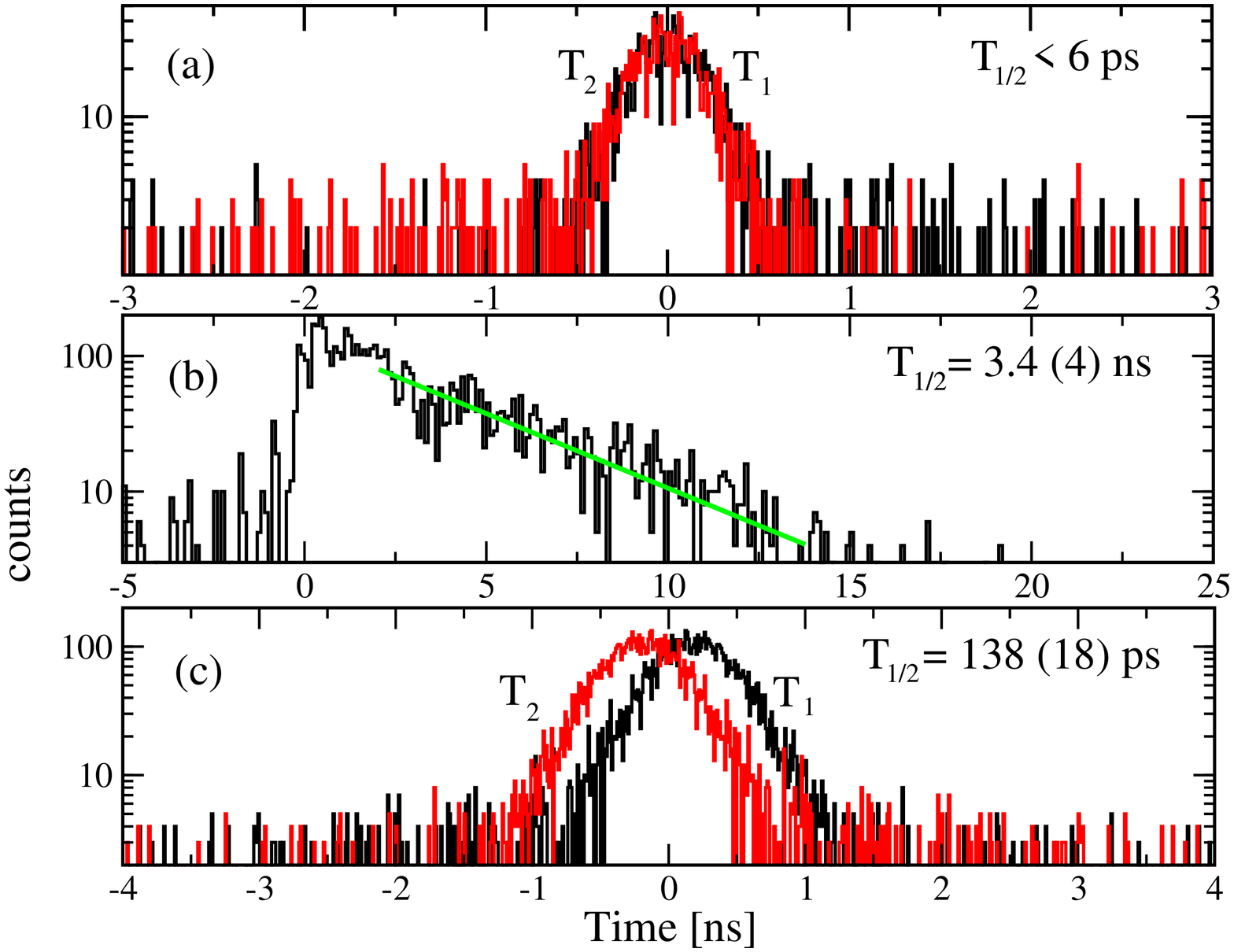}}}
\caption{Time spectra for the 9/2$^{+}$ state (a), for the 19/2$^{+}$ state (b), 
and for the 21/2$^{+}$ state (c) in $^{95}$Mo.} 
\label{time95}
\end{center}
\end{minipage}\hspace{2pc}%
\begin{minipage}{18pc}
\begin{center}
\rotatebox{-90}{\scalebox{0.30}[0.30]{\includegraphics{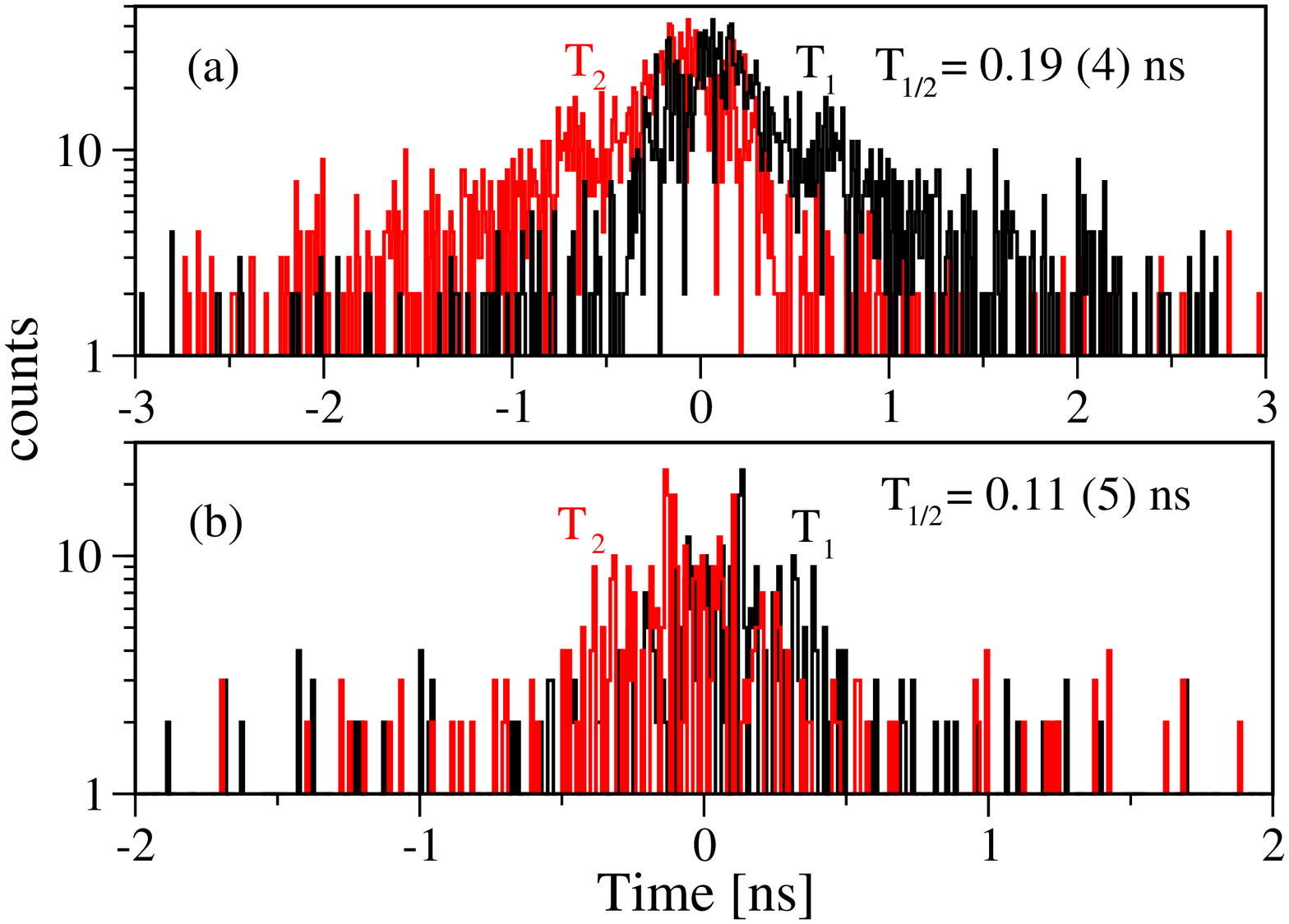}}}
\caption{Time spectra for the 8$_1^{+}$ state (a) and for the 10$_1^{+}$ 
state in $^{96}$Mo.}
\label{time96}
\end{center}
\end{minipage}
\end{figure}

The lowest lying levels in $^{95}$Mo (Fig.~\ref{shf}) have half-lives shorter than 
the electronics time binning of 6 ps/channel. The centroids of the symetric time distributions 
constructed for the $9/2^{+}$ state (Fig. \ref{time95}a) overlap within the time binning, which 
is consistent with the half-life of 2.58(11) ps quoted in NNDC \cite{nds95}.

\begin{table}
\caption{\label{tab}Energy levels and transitions in $^{95}$Mo.}
\begin{center}
\begin{tabular}{cccccccc}
\hline
$J^{\pi\dagger}$&$E_{level}^a$[keV]&$E_{\gamma}^b$ [keV]&$I_{\gamma}$&$\lambda M^{\dagger}$&$T_{1/2}$\\
\hline
5/2$^{+}$   &0.0&&&&stable\\
7/2$^{+}$   &766.6(7)  &766.3&0.80(16) &M1+E2&4.4(7) ps$^{\dagger}$\\
9/2$^{+}$   &948.4(7)  &948.7&2.4(5)   &E2(+M3)&2.58(11) ps$^{\dagger}$\\
11/2$^{+}$  &1542.2(10)&593.6&1&M1+E2  &$\leq$30 ps\\
11/2$^{+}$  &1542.2(10)&775.7&1.09(22) &E2&$\leq$30 ps\\
(13/2$^+$)  &2060.2(17)&1112.0&        & (E2)& \\  
(15/2$^{+}$)&2234.2(15)&173.8&0.086(10)&(M1+E2)&$\leq$30 ps\\
(15/2$^{+}$)&2234.2(15)&692.0&1.39(25) &E2&$\leq$30 ps\\
(17/2$^{+}$)&2582.4(16)&348.1&1.23(11) &M1(+E2) & d\\
(17/2$^{+}$)&2582.4(16)&522.1&0.062(12)&[E2]$^c$& d\\
(19/2$^{+}$)&2620.5(16)&38.1$^\dagger$ &0.056    &[M1+E2]$^c$&3.4(4) ns\\
(19/2$^{+}$)&2620.5(16)&386.4&0.23(2)  &E2&3.4(4) ns\\
(21/2$^{+}$)&2772.4(17)&152.3&1.1(5)   &M1&138(18) ps\\
(25/2$^{+}$)&3675.8(20)&903.4&1.3(3)   &E2& d\\
(29/2$^{+}$)&4143.5(23)&467.7&0.37(5)  &E2& d\\
\hline
\end{tabular}
\end{center}
$^{\dagger}$from NNDC \cite{nds95}, unless otherwise noted;
$^a$from a least-squares fit to E$_\gamma$;
$^b$uncertainty of 1.0 keV;
$^c$from the $J^\pi$ difference;
$^d$no enough statistics in the present experiment.
\end{table}

Fig.~\ref{time95}(b) shows the time distributions, sorted for the $19/2^{+}$ 
state in $^{95}$Mo. The time spectrum is obtained by gating on 152-keV and 386-keV
transitions with the LaBr$_3$:Ce detectors. In addition, to reduce the 
background, gates on 949-keV, 594-keV, 692-keV or 766-keV transitions were 
applied with the HPGe detectors. A half-life of 3.4(4) ns was obtained from the 
slope of the time distribution. According to NNDC, the $19/2^{+}$ state decays via two branches 
with energies of 38 keV and 386 keV. The 38-keV transition is 
highly converted, and its energy is outside of the detectors range of sensitivity. 
Hence, the transition was not observed experimentally in the present experiment.
Therefore, the intensity of the 38-keV transition was estimated from the 
intensity balance performed for the 2582-keV level and after evaluation of the 
levels side feeding. Using the branching ratios obtained from the present data, 
$B(E2)=0.12(9)$ W.u. was obtained.

Fig.~\ref{time95}(c) shows symmetric time spectra for the $21/2^+$ 
state in $^{95}$Mo. The spectrum is gated by the 903-keV and 152-keV transitions
in the LaBr$_3$:Ce detectors. To reduce the background, the same HPGe gates as
for the $19/2^+$ state were imposed. A half-life of 138(18) ps was obtained
from the centroid shift method. 

A weaker reaction channel leads to the even-even $^{96}$Mo nucleus. An analysis 
similar to the one performed for $^{95}$Mo was carried out. Two levels with
picosecond half-lives were observed. Fig.~\ref{time96}(a) shows the time 
distributions obtained for 8$_1^{+}$ state in $^{96}$Mo. The time spectrum is 
gated on 809-keV and 538-keV transitions with LaBr$_{3}$:Ce detectors and on 
812-keV transition with HPGe detectors. A half-life of 0.19(4) ns was 
obtained with the centroid shift method. Hence, the 
$B(E2; 8_1^+\rightarrow 6_1^+)$ is 2.3(5) W.u.
\begin{figure}[t]
\begin{minipage}{18pc}
\begin{center}
\rotatebox{0}{\scalebox{0.30}[0.30]{\includegraphics{Fig5.epsi}}}
\caption{\label{96Molev}Partial level scheme of $^{96}$Mo \cite{Chatt}.}
\end{center}
\end{minipage}\hspace{2pc}%
\begin{minipage}{18pc}
\begin{center}
\rotatebox{-90}{\scalebox{0.30}[0.30]{\includegraphics{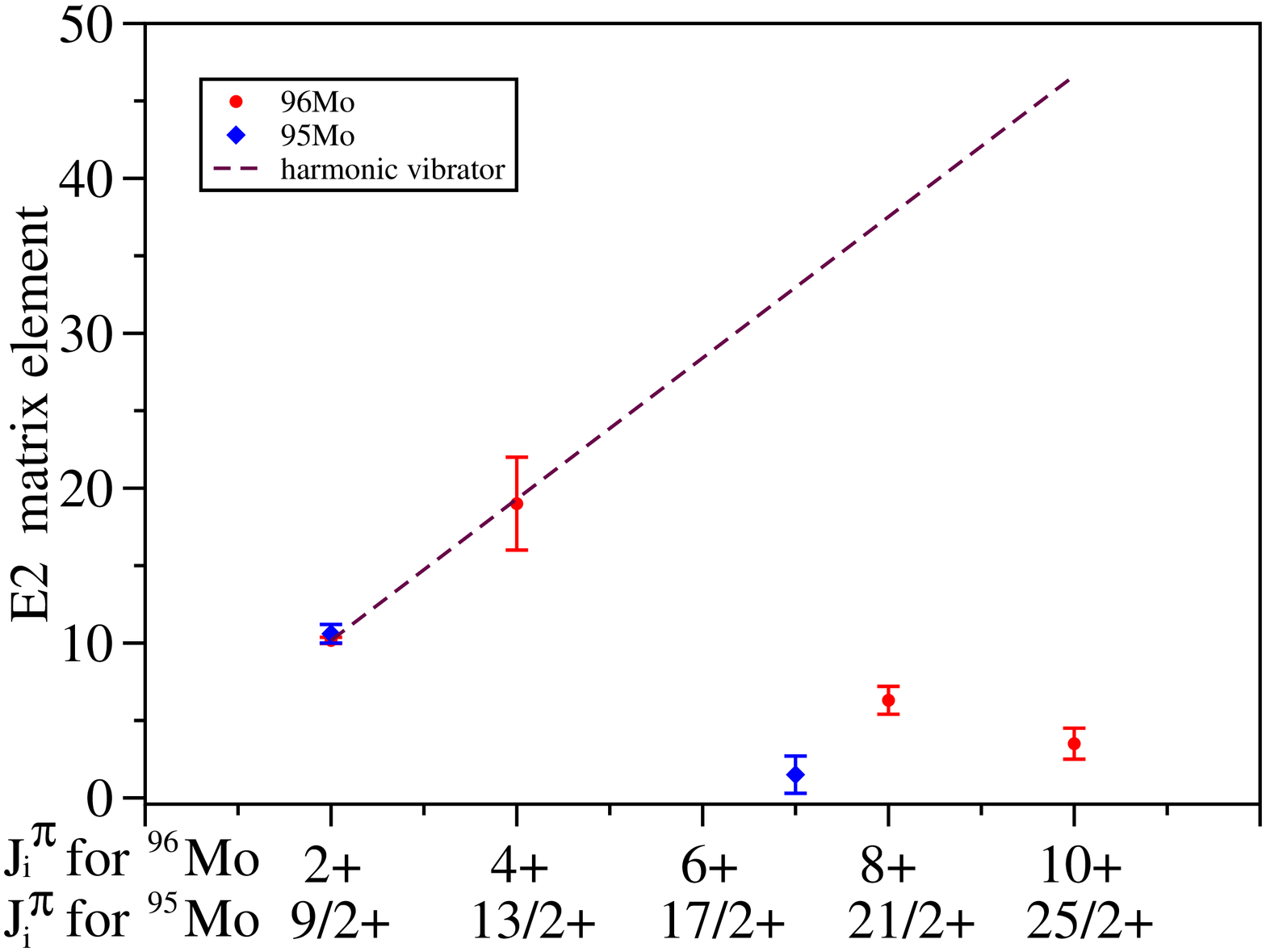}}}
\caption{\label{matrixel}Matrix elements calculated for the $E2$ transitions 
de-exciting the first five yrast states in $^{95}$Mo and $^{96}$Mo and compared 
to the matrix elements, calculated for the harmonic vibrator and normalized with 
respect to the first phonon in $^{96}$Mo.}
\end{center}
\end{minipage}
\end{figure}

Time spectra for the 10$^{+}$ state in $^{96}$Mo are presented on 
Fig.~\ref{time96}(b). The same gate as in the case of the $8_1^+$ state was 
set on the HPGe detectors. Gates on the 1346-keV and 809-keV transitions imposed on 
LaBr$_{3}$:Ce detectors were used to increment the symmetric time spectra. A 
half-life of  0.11(5) ns was obtained using the centroid shift method. The 
transition strength for the 809-keV transition results as 
$B(E2; 10_1^+\rightarrow 8_1^+)=$0.6(3) W.u.

\section{Discussion}
The ground state of $^{95}$Mo is a $J^\pi=5/2^+$ state, which is consistent 
with the
ordering of the neutron single-particle levels in the $A\approx 100$ mass 
region \cite{He94} where $\nu d_{5/2}$ single-particle orbit appears at the 
beginning of the fourth oscillator shell. Next to the $\nu d_{5/2}$
single-particle orbit is  $\nu g_{7/2}$. Indeed, the $J^\pi=7/2^+$ level, 
placed at 767 keV in $^{95}$Mo, decays via a $M1+E2$ transition to the 
ground state with $B(E2;7/2^+\rightarrow 5/2^+)=0.96$ 
W.u \cite{nds95}. The next yrast $J^\pi=9/2^+$ level 
in $^{95}$Mo has an energy of 948 keV, which is close to the $2_1^+$ level
energy of the neighboring even-even nuclei. Also, the $9/2^+$ state decays 
via a strong $E2$ transition with $B(E2)=11.3(6)$ W.u \cite{nds95}. Such a 
state can be interpreted as a $\nu d_{5/2}$ single particle state fully 
aligned with the first $2_1^+$ phonon of an even-even core \cite{Sh61}.
In this respect it is interesting to compare the matrix elements, obtained 
for the $E2$ transitions in the even-even $^{96}$Mo, with the  $E2$ matrix
elements, obtained from the yrast transitions in the even-odd $^{95}$Mo 
nucleus. The matrix elements, calculated from 
$|$$<\psi _f||E2||\psi _i>|=\sqrt{(2J_i+1)\times B(E2; J_i\rightarrow J_f)}$
for the yrast transitions in $^{95,96}$Mo, are shown on Fig.~\ref{matrixel}. 
The experimental matrix elements are also compared to the harmonic vibrator 
matrix elements, which were parameterized from the first excited state in 
$^{96}$Mo. 
The $<2^+||E2||4^+>$ matrix element, calculated for $^{96}$Mo \cite{nds96}, remarkably
coincides with the harmonic vibrator $<2^+||E2||4^+>$ matrix element. Also, 
the $<5/2^+||E2||9/2^+>$ matrix element, calculated from the $^{95}$Mo data,
is consistent with its interpretation as a fully aligned particle-core coupled 
state. Further, with the increase of the angular momentum, the 
experimental matrix elements deviate from the harmonic vibrator description. 
In fact, the experimental matrix elements are even smaller than the  
$<0_1^+||E2||2_1^+>$ matrix element. In this spin range, the 
odd- and even- mass molibdenum matrix elements are again in the same range.
In order to account for the high-$J$ value, these core excited states most 
probably involve high-$j$ single-particle excitations.

\section{Acknowledgments}
This work is supported by the Bulgarian Science Fund - contracts DMU02/1, DRNF02/5, DID-02/16, 
Romanian UEFISCDI - contract IDEI 115/2011 and by a Bulgarian-Romanian partnership contract, 
numbers DNTS-02/21 and 460/PNII~Module~III.

\end{document}